# Toward Trusted and Swift UAV Communication: ISAC-Enabled Dual Identity Mapping

Yanpeng Cui, Zhiyong Feng, Qixun Zhang, Zhiqing Wei, Chenlong Xu, and Ping Zhang

## Abstract

The UAV network has recently emerged as a capable carrier for ubiquitous wireless intelligent communication in the B5G/6G era. Nevertheless, the separation of dual identity raises challenges from the perspective of communication efficiency and security, including tedious communication feedback and malicious Sybil attacks. Meanwhile, thanks to the emerging integrated sensing and communication (ISAC) technology, the sensing ability incorporated in communication advances crucial opportunities for accurately and efficiently mapping identity from dual domains. This tutorial discusses the exciting intersection of ISAC and the future intelligent and efficient UAV network. We first describe the motivation scenario and present the framework of the proposed novel ISAC-enabled dual identity solution. The detailed modules of identity production, mapping, management, and authentication are discussed. By endowing UAVs with an advanced capability: opening their eyes when communicating with each other, we detail three typical applications and the advantages of our proposal. Finally, a series of key enabling techniques, open challenges, and potential solutions for ISAC-enabled dual-domain identity are discussed. This tutorial for the intelligent and efficient UAV network brings new insight on providing dual-domain identity via ISAC technology, with an eye on trusted and swift communication research tailored for the 6G UAV network.

## Introduction

To provide wide coverage and on-demand services for the sixth-generation (6G) mobile communication, the unmanned aerial vehicle (UAV) has been envisioned as a capable carrier for ubiquitous wireless intelligent communication thanks to its appealing characteristics, such as wide coverage with elevated altitude, the ability of on-demand deployment and fast responses[1], and so on. Benefiting from the above merits, the intelligent and efficient UAV network, an innately smart, concise, and ultra-dense heterogeneous network that interconnects all things with extremely low latency and high-speed data transmission, will be a powerful engine for the future intelligent 6G era.

Owing to the shared spectrum and the inherent broadcast nature of the wireless medium, UAV nodes are vulnerable to Sybil attacks, that is, an attacker hacks legitimate nodes and illegally claims to have multiple imaginary identities, called Sybil nodes. This forces the other legitimate nodes to incorrectly perceive that new nodes have joined the network. It yields opportunities for various disruptions, including aggregation, resource allocation, routing decisions, time synchronization, and tampering with reputation systems [2]. The form of Sybil attack includes direct and indirect attacks (i.e., attacks via one hop or multiple hops), simultaneous and non-simultaneous attacks (i.e., introducing all Sybil nodes at once or in different intervals of time), and fabricated and stolen attacks (i.e., steal the existing identities or fabricate the non-existing identities). The UAV communication evidently implicates security and efficiency issues, which presents a contradictory and very interesting challenge for the UAV network. For one thing, it hopes to simplify the verification and remove the tedious feedback to ensure swift and concise communication. For another, it has to ensure the reliability of identity authentication to prevent malicious UAVs from eavesdropping and attacks. Table 1 summarizes the capability of conventional schemes regarding Sybil attack detection, which cannot meet the performance requirement of swift and trusted UAV communication due to the following reasons.

The separation of dual identities leads to untrusted, unreliable, and inefficient issues in sensing and communication. The conventional methods of attack detection and message transmission rely on the information declared by neighbors from the radio domain, including the digital identity (DID) (e.g., IP/MAC address, secret keys) or physical states. That is, UAVs only passively "hear" and have to believe what neighbors "say" from the auditory domain (AD), which generally causes the untrusted and unreliability issue. For one thing, it may be the fake information disguised by the Sybil attacker. For another, even if it comes from a legitimate node, there will also be a large error when obtaining physical features only from AD, for example, the GNSS positioning error may achieve several meters [3]. Evidently, they should also have the capability to proactively observe neighbors' physical identity (PID), for example, relative distance and velocity, to obtain additional sensing information. At the time of writing, state-of-the-art techniques, such as circular scanning millimeter-wave Radar, Lidar, and zoom/wide camera, have been utilized on UAVs to endow active sensing ability, we thus would say it is more like a visual domain (VD) relative to AD. Nevertheless, the DID and PID are

*The authors are with Beijing University of Posts and Telecommunications, China; corresponding author is Zhiyong Feng.*

| Relevant Techniques | Principles | Scalability | Static | Mobile | False positives | Anti attack capability | | | | | | Overhead | | Additional hardware required |
|---|---|---|---|---|---|---|---|---|---|---|---|---|---|---|
| | | | | | | DA | IA | FA | STA | SA | NSA | Communication | Processing | |
| Key-based methods | Symmetric key | × | √ | √ | × | √ | √ | √ | √ | √ | √ | High | High | × |
| | Random key pre-distribution | × | √ | × | × | √ | × | √ | × | √ | √ | High | Medium | × |
| Lightweight methods | Radio resource test | √ | √ | × | × | √ | × | √ | × | √ | × | High | High | √ |
| | RSSI-based | √ | √ | × | √ | √ | × | √ | √ | √ | √ | Low | Low | × |
| | TDOA-based | √ | √ | × | √ | √ | × | √ | √ | √ | √ | High | Medium | √ |
| | Neighbor information | √ | √ | × | √ | √ | × | √ | √ | √ | × | High | Low | × |
| | Mobility-based | × | × | √ | √ | √ | × | √ | √ | √ | × | Medium | Low | × |
| | Energy-based | √ | √ | × | √ | √ | √ | √ | × | √ | √ | Low | Low | × |

Direct Attack (DA); Indirect Attack (IA); Simultaneous Attack (SA); Non-simultaneous Attack (NSA); Fabricated Attack (FA); Stolen Attack (STA).
We define scalability as the applicability of a solution when the scale of the UAV network exceeds a hundred nodes.

**TABLE 1**. A brief summary of the existing identity authentication and Sybil attack defense designs.

still separated, which leads to inefficiency problems. For instance, in a cellular-UAV network, the millimeter-wave beam recovery procedure faces intolerable access latency since the PID of a UAV with a given DID is unknown to the ground base station (BS). Another case is that the emergency communication between UAVs also faces intolerable interaction latency since the DID of neighbors with a given PID is unknown to the transmitter. Therefore, UAVs should have the capability to fully utilize dual identity from the dual domain, in order to meet the critical requirement of trusted, reliable, and efficient communication.

The above requirement evidently implicates the precise mapping of dual identity. However, the separation of communication and sensing functionalities of the conventional methods leads to low efficiency and poor accuracy in obtaining PID. Interestingly, integrated sensing and communication (ISAC), which refers to a new information processing technology sharing information and software/hardware resources, realizes the coordination of sensing and communication (S&C) functionalities [4]. As a result, most of the conventional sensing technologies will fall out of favor except ISAC since it simultaneously achieves both S&C functionalities, and also ensures the accuracy and efficiency of sensing.

Motivated by the above considerations, this article presents a novel ISAC-enabled dual identity solution for trusted and swift UAV communication. To the best of our knowledge, this is the first research that enables UAVs to proactively open their eyes while listening to the voice of others, and make it possible to match what they "hear" and "see" for trusted and swift communication. Based on the ISAC technology, the dual identity obtained from AD and VD are accurately mapped, which removes tedious communication feedback, reduces latency and improves the reliability of sensing and communication. Starting from the motivation scenario, we first introduce the general framework of our proposal, including four modules: identity collector, identity match engine, identity processor and identity authenticator. Then we present three typical applications and the advantages of the proposed ISAC-enabled dual identity solution. In the final section, the key enabling techniques, open challenges, and potential solutions are discussed before concluding this article.

## ISAC-Enabled Dual Identity Solution

Let us start by introducing the motivation scenario and then elaborate on the general framework of the ISAC-enabled dual identity solution.

### Motivation Scenario

The UAV Network could be applied to a host of scenarios shown in Fig. 1. Nevertheless, it tends to be exposed to events that might cause severe communication outages and even crash accidents. For instance, as shown in S-1, the directional gain toward the intended receiver would deteriorate rapidly due to dynamic mmWave channels or obstacles. Besides, the shared use of spectrum and the broadcast nature of wireless communication makes it vulnerable to eavesdropping by malicious users, such as the Sybil attack shown in S-2. In addition, as shown in S-3, the self-organized UAV network faces more collisions than the centralized one since the latency of emergency alerts may exceed the safe limit. As per the security and reliability requirements, trusted and swift communication designs naturally arise in the UAV network.

## Framework of ISAC-Enabled Dual Identity Solution

In order to tackle the above-mentioned issues in the motivation scenario, we offer an ISAC-enabled dual identity solution to remove the tedious communication feedback, reduce beam management latency, and ensure communication security. It exploits one of the evident benefits of ISAC technology, namely the signal used to transmit rich information also brings the receiver's accurate physical features to the transmitter via echo. Based on this, the physical features and digital identities of the UAV nodes can be mapped accurately and thereby brings many benefits, which will be discussed later. Figure 2 demonstrates the framework of the ISAC-enabled dual identity solution, which is mainly composed of four modules, including the production, mapping, management, and authentication of identity.

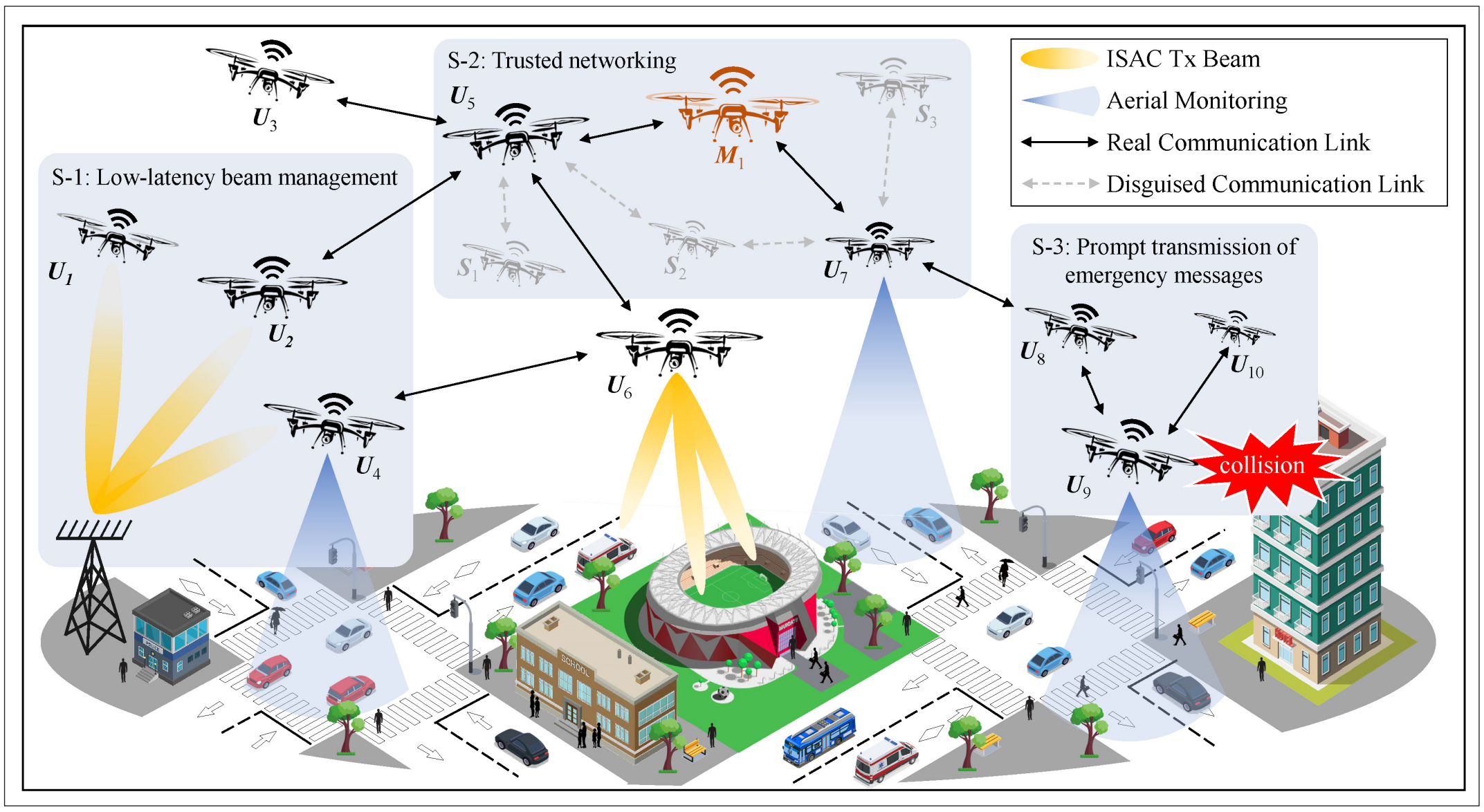

**FIGURE 1.** The motivation scenario and key applications of the intelligent and efficient UAV network. The red, black, and light gray UAVs are malicious nodes, legitimate nodes, and Sybil nodes, respectively. Three typical scenarios S-1, S-2 and S-3 will be discussed at large below.

**Identity Production Module:** The features of UAVs that extract from VD and AD are fed into the production module to produce dual identity, which refers to the physical and digital roles that a transmitter act in the visual and auditory domains of the receiver. The DID represents the label used during the conventional digital authentication process, including the IP/MAC address, unique keys and fingerprint of radio frequency (RF), and so on. The PID denotes the additional feature that characterizes the physical properties of a UAV node, including the wing type, location and mobility, and so on. Based on the reflected echoes of the ISAC signal, different approaches can be exploited to estimate PID. One efficient method is matched-filtering the received echo with a Doppler-shifted and delayed version of the transmitted signal, and the radial velocity and distance can be obtained, then the angle of targets can be estimated via the maximum likelihood estimator. Furthermore, the unique micro-Doppler frequency caused by blade numbers and rotor speeds can be extracted and exploited to distinguish UAVs. In addition, the vision-based methods, for example, you only look once algorithm, can also provide additional measurements of distance, velocity and angle.

It should be emphasized that all the physical features that could distinguish UAVs are used to formulate the PID, which could be obtained from both AD and VD. For one thing, the receiver passively extracts PIDs from the received ISAC signal; for another, it actively observes PIDs via camera or the echo of the transmitted ISAC signal. Whereas DIDs are only obtained from AD since the passively received wireless ISAC signal is the only conveyor for them. The above physical features will be assigned with dynamic weights based on their prevalence [5] to generate PIDs, thus the proposed solution is valid for all types of UAVs even though some of their features are similar. The computational complexity lies in the dynamic similarity calculation, which is O(NM2) in the case where a UAV observes N features from M neighbors.

**Identity Mapping Module:** The PIDs and DIDs obtained from the production module will be transferred to the mapping module to calculate the similarity and perform identity matching. Building upon the similarity of PID pairs from dual domains, it first defines the matching cost Cn(i, j) (e.g., the inverse of the similarity) between the i-th and the j-th identity from VD and AD, respectively. Regarding the matching cost as the edge weight, the weighted bipartite graph-based matching model shown in Fig.2 is established, whose optimization objectives and solutions are adjusted according to performance requirements. For instance, utilizing the Hungarian algorithm with the complexity of O(M3) to minimize the global cost of matching M neighbors' identities. In addition, exploiting the vampire bat optimizer [6] with the complexity of O(MAlog(MC)) to equalize the local cost, where C is the maximum absolute value of competition. A is the number of identity pairs that can be assigned to each other, which is roughly the same as M when eliminating pairs whose similarity is lower than a certain threshold.

**Identity Management Module:** Based on the matching results of the previous module, the management module ensures the accuracy of distinguishing multiple UAV nodes in the dynamic environment. Before the next AD information arrives, the PIDs are only observed in VD and estimated via neural networks or the classical Kalman filter. The identity association techniques, for example, joint probabilistic data association filter and multiple hypothesis tracking, are exploited to associate the identities of multiple UAVs in different time slots. In addition, by minimizing the Euclidean distance between prediction and observation [7], the identity association can be realized more conveniently with the linear complexity. In this way, the accuracy of the subsequent mapping results is ensured.

**Identity Authentication Module:** Building upon the accurate identity management results, the authentication module is a collaboration-based unit to provide trusted local and global views for implementing the above-mentioned applications. Each UAV plays multiple roles, including requestor, witness, and certifier at different times for various purposes. A requestor periodically sends a beacon embedded with information for the purpose of neighbor discovery in AD. It would be received by all the neighboring UAVs within the effective communication range, and they are regarded as the witness of the requestor. The witnesses measure the PIDs of a requestor from dual domains and saves them to the new beacon they are about to send. In this way, the node collects enough PIDs from VD, which correlate to that embedded in the previous beacon received from witnesses. By performing the minimum mean-square error on the PIDs estimated by the certifier and claimed by the requestor, the disguised Sybil identities could be found. For the indirect attack, the local views are merged via the maximum common sub-graph principle to resolve conflicting observations and determine the malicious node, and the Bron-Kerbosch algorithm is one of the efficient methods with the complexity of $\mathcal{O}(3^{n/3})$, where $n$ refers to the number of nodes in the global view [8].

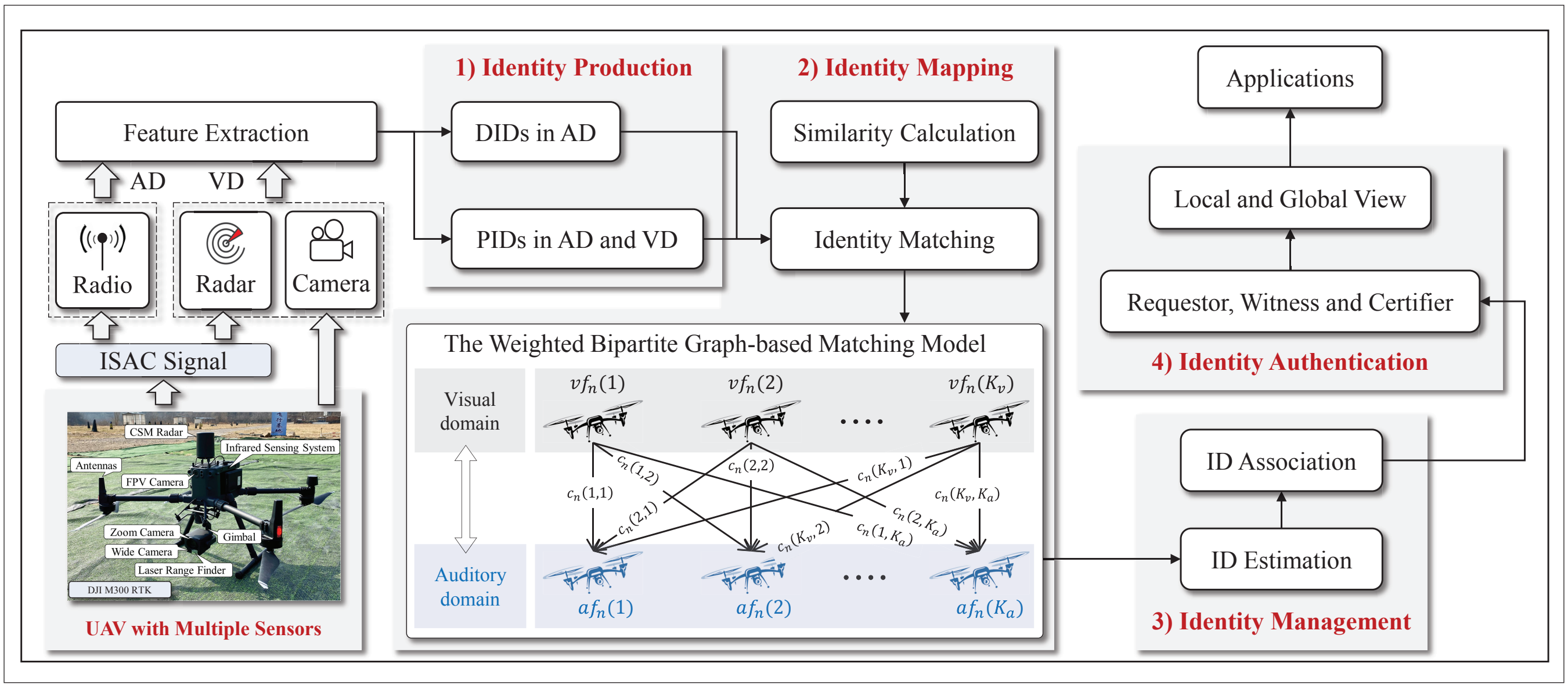

**FIGURE 2**. The framework of the proposed ISAC-enabled dual identity solution. It's mainly composed of four modules, including the production, mapping, management, and authentication of identity.

## Advantages and Potential Applications

This section discusses the advantages of the ISAC-enabled dual identity solution for the UAV network in three applications, including low-latency beam management, the swift transmission of emergency messages, and trusted networking under Sybil attack.

### Low-Latency Beam Management

For the cellular-UAV network shown in Fig. 3a, UAVs play the role of aerial users, and a ground BS that equip with a massive multi-input-multi-output (mMIMO) antenna array serves multiple UAVs simultaneously. It is hoped that the signal-to-noise ratio (SNR) of the user could be improved and the interference to other users could be avoided via beam alignment. 3GPP has specified a set of basic procedures, namely sweeping, measurement determination, and reporting to control the beam management at frequencies above 6 GHz [9]. Nevertheless, due to the highly dynamic topology in 3-D space, the performance requirement of prompt, accurate, and low-overhead tracking of "pencil-like" beams could not be met simultaneously. In addition, the beam search process during the recovery of link failure requires repeated iteration and feedback to determine the best direction, which might miss the fleeting access opportunities for information transmission in high-mobility scenarios. Therefore, the next-generation cellular-UAV network should provide a solution, by which the BS could promptly identify the beam's optimal direction to interconnect UAVs at any given time.

Toward this, the ISAC-enabled dual identity solution shows its first advantage: given the DIDs of UAVs, it provides their corresponding PIDs. In other words, based on the mapping result of DID and PID, the physical characteristics (e.g., distance, azimuth, speed, heading direction) of the UAV with the intended DID could be promptly identified. For the intended receiver with a specific address, parameters (e.g., velocity, angle and distance) that facilitate rapid beam alignment could be directly determined when performing intra-cell beam switching or restoring the failed link. Based on the state measurement via reflected echo of ISAC signal and state estimation and tracking, narrow beams could be directly aligned to the intended UAVs without the iterative search and tedious communication feedback. Here we take the elimination of feedback as an example and illustrate its advantages with the help of real experiments according to the ISAC hardware testbed in our recent work [10]. As shown in Fig. 3c, compared with the feedback-based method, the ISAC technique reduces the total delay by about 4.594 ms and 4.626 ms in the distance of 10 m and 20 m, respectively. In comparison, the unique echo processing delay of about 1 ms is negligible. This strength of ISAC will motivate the dual identity solution for low-latency beam management.

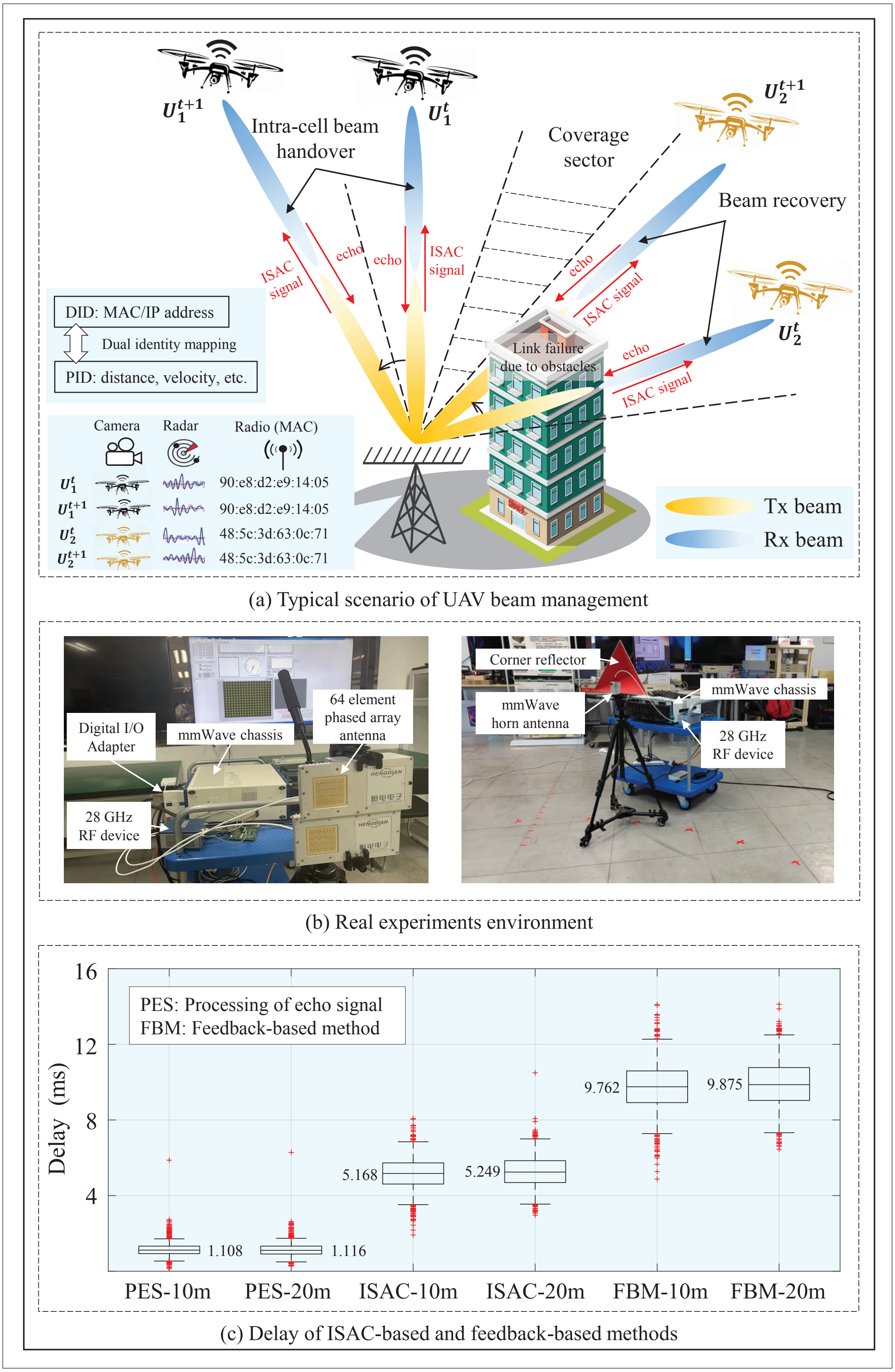

**FIGURE 3**. The real experiments that evaluate the reduction of communication delay of our solution.

## Swift Transmission of Emergency Messages

As shown in Fig. 4a, we consider a scenario where a UAV with a building ahead intends to alert the intended neighbors to avoid potential collisions. The emergency message delivery is conventionally achieved via broadcasting, which generally causes confusion and disturbance at unintended nodes as well as broadcast storm issues in large-scale networks. Although it could be tackled by unicast or multicast to some extent, exchanging beacons is indispensable to confirm the IP address before sending emergency messages, which brings fatal delays.

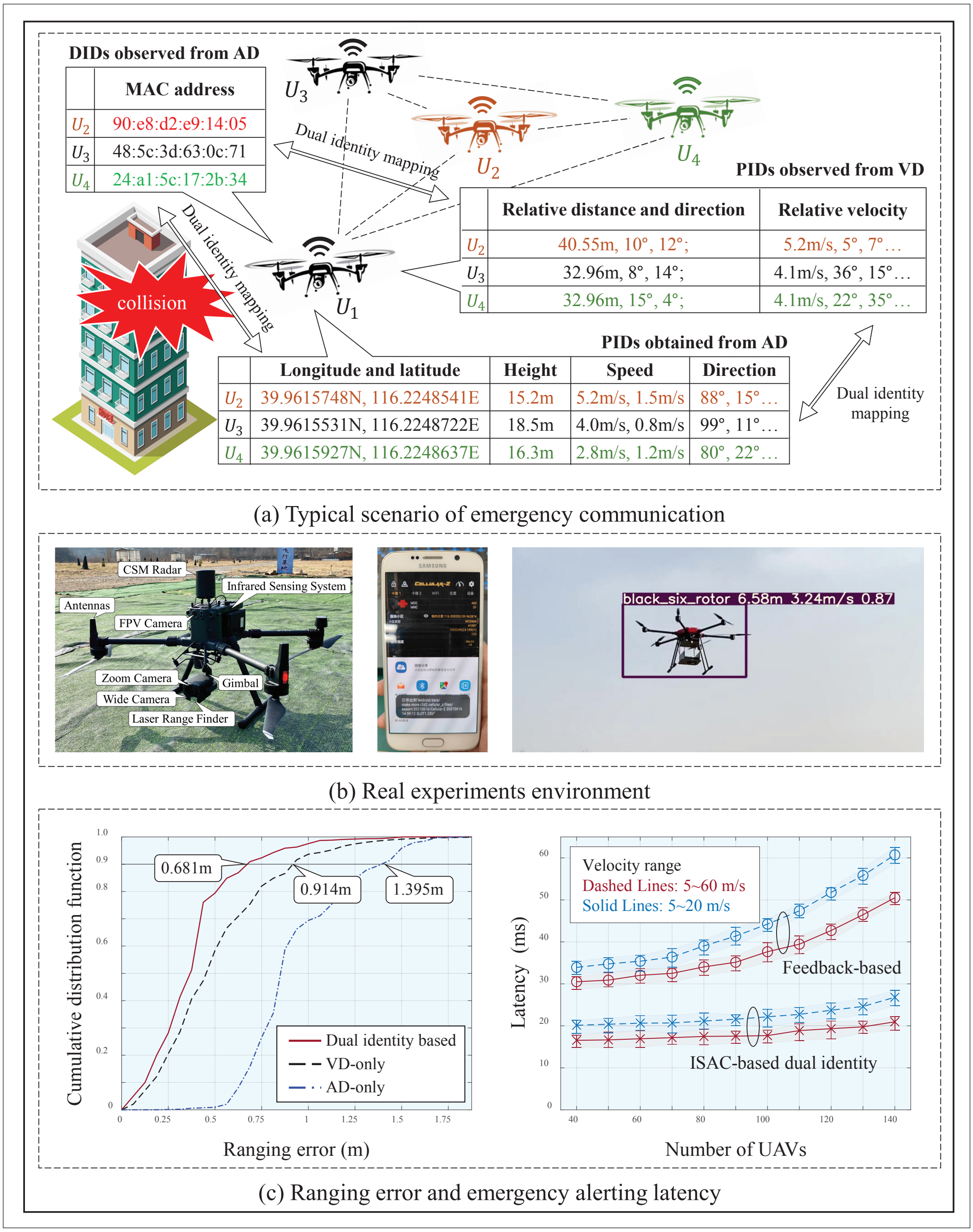

**FIGURE 4**. The real experiments and simulations that evaluate the ranging performance and the latency of emergency alerting of our solution.

Moreover, PIDs are more important than DIDs in this scenario. For instance, $U_1$ would like to alert the neighbor in the most dangerous situation (i.e., $U_2$ rather than $U_3$ and $U_4$ since $U_2$ is the nearest one that is approaching) regardless of its IP address. Nevertheless, the conventional PID acquisition method requires UAVs to periodically exchange beacons embedded with features, which generally causes large overhead and latency since the PID is passively heard from AD. Therefore, in order to accurately and quickly send emergency messages to the intended node, UAVs should have the ability to match the DID and PID obtained from dual domains.

As per the above high-reliability and low-latency requirements in the emergency alerting scenario, the ISAC-enabled dual identity solution manifests its second advantage: given the UAV's PID, it provides the corresponding DID. Specifically, the PID information extracted via the echo of the ISAC signal is associated with the information recently received from AD, thereby realizing the mapping between PID pairs obtained from AD and VD, which reduces the inaccuracy of observing PID from a single domain. Furthermore, DIDs will be coupled with PIDs, and thus the IP/MAC address of the intended UAV with the expected physical features are confirmed promptly. As a result, emergency messages can be sent directly to the intended UAV accurately without tedious beacon exchange and DID confirmation. Here we illustrate the two advantages with the help of real experiments and simulations according to the UAV-based testbed in our recent work [5]. As shown in

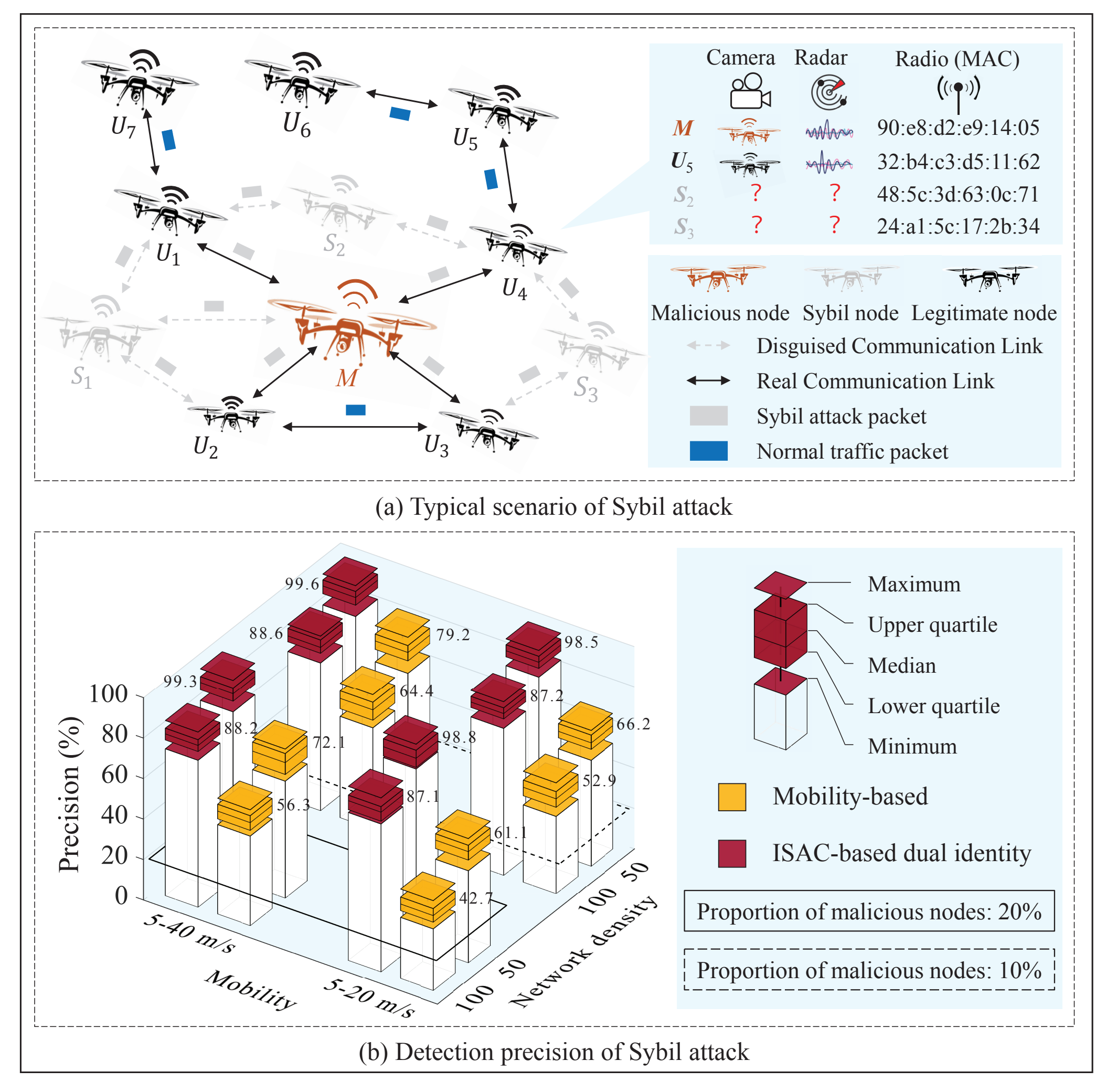

**FIGURE 5**. The simulations that evaluate the precision of Sybil attack detection of our solution.

Fig. 4c, since mapping identities from dual domains significantly reduce individual inaccuracy and unreliability, a 90 percentile ranging error of 0.681 m is achieved by our solution, which is lower than AD-only and VD-only ranging methods by 51.18 percent and 25.49 percent, respectively. In addition, compared with the feedback-based scheme, our solution reduces the emergency alerting latency by 66.54 percent on average due to the accurate mapping of dual identity, which significantly improves the efficiency and safety of the UAV network.

## Trusted Networking under Sybil Attack

The sensor-rich and ISAC-enabled UAV network can easily leak out private information, and even become a dangerous bomb in the sky once hacked. As shown in Fig. 5a, we consider a Sybil attack scenario where a malicious UAV M illegally generates Sybil nodes $S_1 \sim S_3$ to disrupt legitimate UAV $U_1 \sim U_4$. It claims to be different from $M$ in various dimensions, including location, speed, digital identity, and so on. Therefore, one of the legitimate UAVs, for example, $U_4$, will discover four different neighbors with various identities and physical features from AD. As mentioned earlier, it is not trivial to detect the Sybil attack only from AD. Fortunately, since two of them are in disguise, namely they don't exist in reality, it provides us an opportunity to recognize them from VD. Therefore, how to accurately and actively observe the physical features of neighbors from VD is the premise of detecting the Sybil attack accurately. Among all the sensing methods, we believe the ISAC technology is the most reliable one, and the specific reason will be discussed later. However, the information obtained from VD only provides us with the physical features of two real nodes rather than their DIDs. In other words, we are only aware that two of them are disguised, but it's difficult to tell which belongs to the Sybil nodes.

To tackle the above security issues in the Sybil attack scenario, the ISAC-enabled dual identity solution presents its third advantage: given the accurate mapping results of dual identities from AD and VD, the malicious node together with its Sybil identity could be detected successfully. Since there is a large deviation between the disguised identities that M claims in AD and the results that $U_4$ observes in VD, the Sybil nodes could be easily detected by matching the PID pairs via similarity. Besides, the malicious node is found based on the maximum common sub-graph, which could be drawn based on the information shared by legitimate UAVs. It avoids the drawbacks of the AD-only method presented in Table 1, including the strict time synchronization, no support for mobile net-works, high complexity, and large latency, also it solves the difficulty of VD-only methods in determining the DID of malicious nodes. Here we illustrate the advantages with the help of simulations according to the environment in our recent work [5]. As shown in Fig. 5b, thanks to the accurate sensing and mapping results of dual identity, our solution outperforms the mobility-based scheme in detection precision by 31.55 percent on average.

## Key Enabling Techniques, Open Challenges and Solutions

This section discusses the key enabling technologies along with the unique open challenges and the potential solutions of the ISAC-enabled dual identity.

### ISAC

Generally, UAVs are repeatedly switched between the long-term connected mode and the temporary idle mode. For the connected mode, radio signals are frequently transmitted and received while for the idle mode, there is no communication demand temporarily. In the connected mode, where the performance of both communication and sensing are jointly considered, most of the conventional sensing technologies will fall out of favor except ISAC since it utilizes the unified RF transceivers and frequency resources to achieve both S&C functionalities. ISAC transmitter is able to perceive the angle of arrival of the echo signal reflected from the intended receivers and infer their distance and velocity according to the time delay and Doppler shift. It also outperforms the communication-only feedback-based protocols and the other sensing method in the following aspects:

- It removes the pilot and the consequent tedious feedback for the communication-based PID sensing, thus the PID can be obtained efficiently.
- It realizes the continuous angle estimation without feedback quantization, which brings a high confidence feature to identity production.
- The use of the entire echo signal block for radar sensing brings significant matched-filtering gain, and the estimation accuracy of relative distance and radial velocity is improved accordingly, providing identity mapping with precise PIDs.
- It actively senses the neighbors' DID from VD while transmitting DID and PID in AD, which is more efficient than any other AD-only sensing method for mapping dual identity.

The above benefits of the ISAC technique provide accurate PIDs with the lowest overhead and ensure the performance of the dual identity solution for the UAV network, which is essential in the applications of low-latency beam management, the swift transmission of emergency messages, and trusted networking under Sybil attack.

**Challenges:** The data frames embedded in the ISAC signals make the communication vulnerable to eavesdropping by malicious UAVs, which imperil data privacy as well as the user's identity. This yields a challenging trade-off for UAVs. On one hand, it hopes to concentrate power toward the intended receiver, while on other hand, the power should be limited to avoid eavesdropping [11]. Besides, most of the existing ISAC schemes are designed for 2-D scenarios, for example, vehicle-to-infrastructure [7], while the 3-D ISAC design is rare. Thus it's a challenge to implement the ISAC-based dual identity for UAV beam tracking. The measurement of PIDs in 3-D is generally performed in a polar coordinate system while the state evolution model is established in a Cartesian one, and the conversion error generally leads to biased estimation and degrades filtering performance. Additionally, the narrow beam is hard to directly aligned to UAVs since the echo signal only brings the PID information from the last epoch.

**Potential Solutions:** Information security could be ensured by prudently and dynamically suppressing the signal-to-interference-noise ratio (SINR) at the target receiver, which is equivalent to limiting the achievable rate. When the location of the intended receiver is perfectly perceived, a narrow beam pattern should be implemented to improve the anti-eavesdrop performance, thereby achieving a high level of secrecy rate. On the contrary, a wider beam pattern should be formed to guide the same power over the possible area while reducing the power of the main beam. Employing artificial noise at the transmitter is another effective solution to minimize the SNR received at radar targets while maintaining the SINR required by legitimate users [12]. Besides, different from the 2-D scene based on uniform linear array, the design of beamforming matrix and steering vector and the processing model of the echo signal in the 3-D scene should be designed based on uniform planar array. It should be emphasized that the BS should perform one- and two-step predicting when performing beam tracking, that is, formulating transmit beams by using the one-step predictions and then sending a signal that contains the information of the two-step predictions to avoid the one-step prediction angle becoming outdated at the next epoch. As a result, UAVs will correspondingly formulate receive beamformer at the next epoch based on the two-step prediction of the angle parameter. In addition, exploiting the modified unbiased converted measurements Kalman filter will be an effective solution to ensure measurement and estimation accuracy in 3-D scenarios.

### Multi-Sensory Integration

Considering that the ISAC technique is invalid in idle mode since no signal needs to be transmitted, the PID needs to be measured and tracked in the identity management module by other methods. The fusion of Radar, camera, and Lidar is crucial when UAVs are in idle mode, and the ISAC can be suspended since one of its functionalities, namely communication, is unnecessary. Moreover, multimodal sensory data typically have varying confidence levels when accomplishing different tasks [13]. For instance, vision-based sensing has an excellent performance in object classification, while the ranging ability of radio is better than that of a camera. Besides, the sensing resolution of Lidar is generally higher than that of Radar and camera, and optical perception is sensitive to the weather while Radar sensing is more robust in various environments. Therefore, it would be necessary to perform multi-sensory fusion to integrate the sensing information, which could significantly reduce individual inaccuracy and unreliability while improving the overall performance [14].

**Challenges:** The camera or Lidar are available in real-time while ISAC signals are transmitted and received at relatively large time intervals, resulting in fewer exchanges of AD information. The PID observed from VD has a higher refresh rate than that obtained from AD, which may cause poor mapping performance since the matching of PID and The camera or Lidar are available in real-time while ISAC signals are transmitted and received at relatively large time intervals, resulting in fewer exchanges of AD information. The PID observed from VD has a higher refresh rate than that obtained from AD, which may cause poor mapping performance since the matching of PID and DID cannot be performed at anytime, namely the asynchrony issue.

DID cannot be performed at anytime, namely the asynchrony issue. In addition, owing to the different sensing ranges of AD and VD, UAVs would "see" but not "hear" others and vice versa, and the Sybil attack also makes the number of nodes observed in dual domains unequal (i.e., the asymmetry issue), which causes difficulty in matching identity.

**Potential Solutions:** To tackle the asynchrony problem, the PID observed from VD should be constantly tracked and estimated until the next AD information is available, and the multi-UAVs can be distinguished by operating correlation gates and point-track association. By augmenting the asymmetric cost matrix, that is, establishing virtual identities in the deficient domain and setting their matching cost to the identities in the completed domain as infinity, the asymmetry issue could be solved without introducing additional complexity [6]. In addition, the rate and distribution of topology change arrival of UAV network could be analyzed based on queuing theory [15]. As a result, no topology changes will be missed in AD, thus bringing significant reliability for mapping dual identity.

## Conclusion

This article discusses the exciting intersection of ISAC and the future intelligent and efficient UAV network. Starting from the security, latency, and reliability requirements of UAV networks, we first introduce the motivation scenarios. Then we present the framework and detailed modules of the proposed ISAC-enabled dual identity solution that can realize trusted and swift communication. We focus on three key applications in typical scenarios of UAV networks and explain how our proposal solves the crucial issues. Finally, key enabling technologies, along with technical challenges and corresponding potential solutions are discussed. This tutorial on intelligent and efficient UAV networks offers a novel idea to enhance trusted and swift communication performance via ISAC-enabled dual-domain identities, which holds the promise of inspiring exciting research for 6G UAV networks.

## Acknowledgment

This work was partly supported by National Key Research and Development Project (2020YFA0711300), National Natural Science Foundation of China (62022020, 61790553), and BUPT Excellent Ph.D. Students Foundation (CX2022208).

## Biographies

Yanpeng Cui [S'20] received the B.S. degree from the Henan University of Technology, Zhengzhou, China, in 2016, and the M.S. degree from the Xi'an University of Posts and Telecommunications, Xi'an, China, in 2020. He is currently pursuing the Ph.D. degree with the School of Information and Communication Engineering, Beijing University of Posts and Telecommunications (BUPT), Beijing, China. His current research interests include the Flying ad hoc networks as well as integrated sensing and communication for UAV networks.

Zhiyong Feng [M'08, SM'15] received her B.S., M.S., and Ph.D. degrees from BUPT, Beijing, China. She is a Professor with the School of Information and Communication Engineering, BUPT, and the director of the Key Laboratory of Universal Wireless Communications, Ministry of Education, China. Her research interests include wireless network architecture design and radio resource management in fifth generation (5G) mobile networks, spectrum sensing and dynamic spectrum management in cognitive wireless networks, universal signal detection and identification, and network information theory. She is active in standards development, such as ITU-R WP5A/5C/5D, IEEE 1900, ETSI, and CCSA.

Qixun Zhang [M'12] received the B.E. and the Ph.D. degree from BUPT, Beijing, China, in 2006 and 2011, respectively. From March to June 2006, he was a Visiting Scholar at the University of Maryland, College Park, Maryland. From Nov. 2018 to Nov. 2019, he was a Visiting Scholar in the Electrical and Computer Engineering Department at the University of Houston, Texas. He is a Professor with the Key Laboratory of Universal Wireless Communications, Ministry of Education, and the School of Information and Communication Engineering, BUPT. His research interests include 5G mobile communication system, integrated sensing and communication for autonomous driving vehicle, mmWave communication system, and unmanned aerial vehicles (UAVs) communication. He is active in ITU-R WP5A/5C/5D standards.

Zhiqing Wei [S'12, M'15] received his B.E. and Ph.D. degrees from BUPT in 2010 and 2015. Now he is an associate professor at BUPT. He was granted the Exemplary Reviewer of IEEE Wireless Communications. Letters in 2017, the Best Paper Award of Int'l. Conf. Wireless Communications and Signal Processing 2018. He was the Registration Co-Chair of IEEE/CIC Int'l. Conf. Communications in China (ICCC) 2018 and the publication Co-Chair of IEEE/CIC ICCC 2019. His research interest is the performance analysis and optimization of mobile ad hoc networks.

Chenlong Xu received the B.E. degree from BUPT, Beijing, China, in 2021. He is currently pursuing the M.E. degree with the School of Information and Communication Engineering, BUPT, Beijing, China. His current research interests include radar detection as well as integrated sensing and communication for UAV network.

Ping Zhang [M'07, SM'15, F'18 received his M.S. degree in electrical engineering from Northwestern Polytechnical University, Xi'an, China, in 1986, and his Ph.D. degree in electric circuits and systems from BUPT, Beijing, China, in 1990. He is currently a Professor with BUPT, a Professor with the School of Information and Communication Engineering, Beijing University of Posts and Telecommunications (BUPT), and the Director of the State Key Laboratory of Networking and Switching Technology. He is also an Academician with the Chinese Academy of Engineering (CAE). His research interests mainly focus on wireless communications. He is also a member of the IMT-2020 (5G) Experts Panel and the Experts Panel for China's 6G development. He served as the Chief Scientist for the National Basic Research Program (973 Program), an Expert for the Information Technology Division of the National High-Tech Research and Development Program (863 Program), and a member of the Consultant Committee on International Cooperation, National Natural Science Foundation of China.